\documentclass[12pt,english,aps,manuscript]{article}
\usepackage[T1]{fontenc}
\usepackage[latin1]{inputenc}
\usepackage{geometry}
\usepackage{amsmath}
\usepackage{color}
\usepackage{amssymb}
\usepackage[numbers]{natbib}

\makeatletter
\newcommand{\bee}{\begin{equation}}
\newcommand{\ee}{\end{equation}}
\newcommand{\beea}{\begin{eqnarray}}
\newcommand{\eea}{\end{eqnarray}}

\usepackage{babel}
\makeatother

\begin{document}

\section*{\textmd{\normalsize COLO-HEP-549}}

\begin{center}
\textbf{\Large Classical and Quantum SUSY Breaking Effects in IIB
Local Models}
\par\end{center}{\Large \par}

\begin{center}
{\large S. P. de Alwis$^{\dagger}$} 
\par\end{center}

\begin{center}
Physics Department, University of Colorado, \\
 Boulder, CO 80309 USA 
\par\end{center}

\begin{center}
\vspace{0.3cm}
\par\end{center}

\begin{center}
\textbf{Abstract}
\par\end{center}

\begin{center}
\vspace{0.3cm}
\par\end{center}

We discuss the calculation of soft supersymmetry breaking terms in
type IIB string theoretic models in the Large Volume Scenario (LVS).
The suppression of FCNC gives a lower bound on the size of the compactification
volume. This leads to soft terms which are strongly suppressed relative
to the gravitino mass so that the dominant contribution to the gaugino
masses comes from the Weyl anomaly. The other soft terms are essentially
generated by the renormalization group running from the string scale
to the TeV scale.

\vfill

$^{\dagger}$ dealwiss@colorado.edu

\eject

\section{Introduction}

Theories of supersymmetry (SUSY) breaking and transmission from a
hidden to a visible sector has been the subject of much discussion
over the last two to three decades. Much of this discussion has had
little to do with string theory and often it has been conducted purely
within a global SUSY framework. However a theory of supersymmetry
breaking must necessarily be embedded within a (${\cal N}=1$) supergravity
(SUGRA) which is derived from string theory. The following is a summary
of the arguments leading to this assertion.

\begin{enumerate}
\item \textcolor{black}{Adding a set of explicit soft SUSY breaking terms
to a global theory (like the Minimally Supersymmetric Standard Model
(MSSM)) leads to far too much arbitrariness - it does not give us
a theory.}
\item Spontaneous SUSY breaking in Global SUSY leads to a cosmological constant
(CC) at the SUSY breaking scale which cannot be fine-tuned to zero.
\item \textcolor{black}{A theory of SUSY breaking is therefore necessarily
a SUGRA with a scalar potential which has a minimum that breaks SUSY
spontaneously.}
\item A SUGRA needs to be embedded in string theory in order to have a quantum
mechanically consistent and complete theory. 
\end{enumerate}
\textcolor{black}{The main problem in relating string theory to phenomenology
is that the starting point of the theory is in ten dimensions. While
there are only five weakly coupled string theories (which are related
to each other through various dualities) the number of four dimensional
`compactifications' is extremely large. At the time of the second
string revolution of the mid eighties, it was hoped that in spite
of the existence of many compactifications, the number of theories
with stabilized moduli (the fields governing the size and shape of
the compact manifold) is small if not just one. However it was realized
through the work of many authors (for a review see \citep{Grana:2005jc,Douglas:2006es})
culminating in that of \citep{Giddings:2001yu} (GKP) and \citep{Kachru:2003aw}
(KKLT) that the number of such four dimensional models is extremely
large. Thus at the current stage any discussion of the phenomenological
consequences of string theory must proceed by first imposing a set
of experimental inputs (in addition to requiring a compactification
to four dimensional ${\cal N}=1$ supergravity). These are: }

\begin{itemize}
\item \textcolor{black}{CC is tiny $\sim O((10^{-3}eV)^{4})$ }
\item \textcolor{black}{No light scalars with gravitational strength coupling}
\item \textcolor{black}{SUSY partner masses $\gtrsim O(100GeV)$ }
\item \textcolor{black}{Lightest Higgs $>114GeV$}
\item \textcolor{black}{Flavor changing neutral currents (FCNC) suppressed}
\item \textcolor{black}{No large CP violating phases}
\end{itemize}
\textcolor{black}{The first of these is achieved by ensuring that
there is a sufficiently large number of flux configurations such that
there would be many solutions that realize this value. For typical
Calabi-Yau compactifications of IIB string theory this certainly is
the case. The second is achieved by a combination of fluxes and non-perturbative
(NP) effects. The remaining four constraints are dependent on the
particular mechanism of supersymmetry breaking and transmission. }

In this paper we will discuss a theory of supersymmetry breaking that
emerges from the so-called Large Volume Scenario (LVS) of type IIB
string compactifications \citep{Balasubramanian:2005zx}. In addition
to fluxes which stabilize the dilaton and complex structure moduli,
as in the original KKLT model \citep{Kachru:2003aw} non-perturbative
(NP) effects may be used to stabilize the Kaehler moduli. Although
there appears to have been some controversy about the latter in the
literature the issue seems to have been settled - at least when the
cycles in question are not wrapped by branes that support a gauge
theory with chiral fermions. (For a recent comprehensive discussion
and for references to earlier work see \textcolor{black}{\citep{Berglund:2005dm}}).

However for a four-cycle which is wrapped by D7 branes carrying a
chiral gauge theory (such as the MSSM) the situation appears to be
different. In this case it has been argued in \citep{Blumenhagen:2007sm}
that the chirality (of the MSSM) precludes the stabilization of the
four cycle which they wrap by NP effects. The argument depends on
the observation that in a D-brane construction of chiral theory there
would be an anomalous $U(1)$ gauge group, which in effect requires
the presence of charged matter field factors in the NP superpotential
contribution that depends on the relevant four cycle volume. It has
been argued in \citep{Blumenhagen:2007sm} (see also \citep{Conlon:2008wa,Blumenhagen:2009gk})
that such matter fields must have zero vacuum values, so that effectively
this contribution would be absent. This means also that this cycle
would shrink below the string scale (this follows from examining the
associated D-term potential). Effectively the situation becomes similar
to that of having a D3 brane at a singularity (for more discussion
see below). It is not clear to the author that this argument has been
rigorously established, however it appears that requiring a reasonable
phenomenology (in particular that MSSM fields should not acquire vaccum
values at the scales at which the moduli are stabilized) seems to
justify such a scenario. In any case we will take the attitude that
such an outcome yields an interesting set up whose phenomenology is
worth investigating. 

After discussing the basic physical inputs and reviewing the pertinent
results of \citep{Conlon:2008wa} and \citep{Blumenhagen:2009gk},
we go on to discuss the classical soft terms that arise from this
class of theories (the details of the calculations are given in Appendix
A). In particular we will find that the classical soft mass squared
is positive definite. This and the classical CC are both highly suppressed
by a power of the (large) volume. Furthermore we show that the suppression
of FCNC effects imply that there is a lower bound on the volume. If
we ignore Weyl anomaly effects then comparison of the classical FCNC
effects with the flavor diagonal classical masses, leads to a large
volume ${\cal V}\gtrsim10^{12}$ in Planck units. 

However Weyl anomaly effects (usually called AMSB) changes the phenomenology
of this class of theories. As shown in Appendix B the Weyl anomaly
gives an additional set of terms (calculated by Kaplunovsky and Louis
(KL) \citep{Kaplunovsky:1994fg}) to the gauge coupling function,
leading to a contribution  to the gaugino mass that is much larger
than the classical one. This in turn drives the scalar masses by the
mechanism of gaugino mediation \citep{Kaplan:1999ac,Chacko:1999mi}.
However the lower bound on the CYO volume implies a tension between
having TeV scale soft masses (to address the hierarchy problem) and
making the sGoldstino heavy enough to avoid the cosmological modulus
problem.

\section{Generalities}

We follow the notation and discussion of \citep{Conlon:2008wa} and
\citep{Blumenhagen:2009gk}. We also set $M_{P}\equiv(8\pi G_{N})^{-1/2}=2.4\times10^{18}GeV=1$.

The superpotential, Kaehler potential and gauge kinetic function for
the theory under discussion are,

\begin{eqnarray}
W & = & \hat{W}(\Phi)+\mu(\Phi)H_{1}H_{2}+\frac{1}{6}Y_{\alpha\beta\gamma}(\Phi)C^{\alpha}C^{\beta}C^{\gamma}+\ldots,\label{eq:W}\\
K & = & \hat{K}(\Phi,\bar{\Phi})+\tilde{K}_{\alpha\bar{\beta}}(\Phi,\bar{\Phi})C^{\alpha}C^{\bar{\beta}}+[Z(\Phi,\bar{\Phi})H_{1}H_{2}+h.c.]+\ldots\label{eq:K}\\
f_{a} & = & f_{a}(\Phi).\label{eq:f_a}\end{eqnarray}
Here $\Phi=\{\Phi^{A}\}$ and $C^{\alpha}$ are chiral superfields
(including the two Higgs doublets $H_{1,2}$) that correspond to the
moduli and MSSM/GUT fields respectively. Also\begin{eqnarray}
\hat{K} & = & -2\ln\left({\cal V}+\frac{\xi}{2}\left(\frac{(S+\bar{S})}{2}\right)\right)-\ln\left(i\int\Omega\wedge\bar{\Omega}(U,\bar{U})\right)-\ln(S+\bar{S}),\label{eq:hatK}\\
\hat{W} & = & \int G_{3}\wedge\Omega+\sum_{i}A_{i}e^{-a_{i}T^{i}}.\label{eq:hatW}\end{eqnarray}
Here ${\cal V}$ is the volume (in Einstein frame) of the internal
manifold and the $\xi$= -($\chi\zeta(3)/2(2\pi)^{3}$) term is a
correction term that is higher order in the $\alpha'$ expansion.
For typical Calabi-Yau manifolds $\xi\sim O(1)$. $S$ is the axio-dilaton,
$U=\{U^{m}\}$ represents the set of ($m=1,\ldots,h_{21}$) complex
structure moduli and $T^{i}$ ($i=1,\ldots,h_{11}$) are the (complexified)
Kaehler moduli. The type of Calabi-Yau manifolds that we consider
are of the `Swiss cheese' type. In the simplest such manifold consistent
with our requirements the volume may be written as%
\footnote{In order to simplify the notation we have rescaled the moduli by replacing
those used in \citep{Blumenhagen:2009gk} by $\tau_{i}\rightarrow\eta_{i}^{-1}\tau_{i}$
. Correspondingly we have also replaced $a_{i}\rightarrow\eta_{i}a_{i}$.%
}\begin{equation}
{\cal V}=\tau_{b}^{3/2}-\tau_{s}^{3/2}-\tau_{a}^{3/2}.\label{eq:Swisscheese}\end{equation}
In the above the tau's are Kaehler moduli which control the volume
of the four cycles with $\tau_{b}$ effectively determining the overall
size of the CY. While in explicit calculations in the the rest of
the paper, we will use \eqref{eq:Swisscheese} for the sake of simplicity
it should be clear from the discussion that the results would hold
even in a more general CY manifold of this type. 

It turns out that in order to realize the Large Volume Scenario (LVS)
with the MSSM located either on a $D3$ brane at a singularity or
a seven-brane wrapping a four cycle on a CY orientifold, the total
number of 2/4 cycles $h_{11}\ge3$. At least one of these cycles ($\tau_{s}$
in the above) must either be wrapped by a Euclidean D3 instanton or
by a stack of seven branes with a condensing gauge group that will
generate non-perturbative effects. As pointed out in \citep{Blumenhagen:2007sm}
the MSSM cannot be located on this cycle - hence the need for (at
least) one more cycle ($\tau_{a}$). It has been argued in \citep{Conlon:2008wa,Blumenhagen:2009gk}
that this cycle shrinks to zero unless stabilized at the string scale
by non-perturbative string effects. In any case the F/D term associated
with the corresponding Kaehler modulus is zero. Let us briefly review
this argument.

The potential for the moduli is (assuming that the minimum would be
at large ${\cal {\cal {\cal V}}}$ and expanding in it) \begin{eqnarray}
V & = & V_{F}+V_{D}.\label{eq:pot1}\\
V_{F} & = & \frac{4}{3}g_{s}(a|A|)^{2}\frac{\sqrt{\tau_{s}}e^{-2a\tau_{s}}}{{\cal V}}-2g_{s}a|AW_{0}|\frac{\tau_{s}e^{-a\tau_{s}}}{{\cal V}}+\frac{3}{8}\frac{\xi|W_{0}|^{2}}{g_{s}^{1/2}{\cal V}^{3}}+\ldots,\label{eq:VF}\\
V_{D} & = & \frac{f}{2}D^{2},\, D=f^{-1}k^{i}K_{i}.\label{eq:VD}\end{eqnarray}
In the above we've used the absence of a non-perturbative superpotential
for the modulus of the MSSM cycle and the fact that the effective
axionic partner of $\tau_{s}$ is stabilized at an odd multiple of
$\pi$ (giving the sign flip of the second term of $V_{F}$). The
phases of $A$ and $W_{0}$ (the flux superpotential) can then be
set to zero without loss of generality \citep{Balasubramanian:2005zx}.
The D-term comes from the anomalous $U(1)$ (with Killing vector field
$k$) living on the MSSM cycle under which the standard model fields
and the modulus $\tau_{a}$ are charged and we have set the matter
fields which are charged under this $U(1)$ to zero following the
arguments of \citep{Blumenhagen:2007sm} \citep{Conlon:2008wa,Blumenhagen:2009gk}.
The $U(1)$ gauge coupling function is linear in $\tau_{a}$ and $S$.
Also one has $K_{a}\sim\tau_{a}^{\alpha}/{\cal V},\,\alpha>0$, and
$D\propto1/{\cal V}$. The F-term potential is minimized at\begin{eqnarray}
e^{-a\tau_{s}} & \simeq & \frac{3}{4}\frac{W_{0}}{aA{\cal V}}\sqrt{\tau_{s}}\left(1-\frac{3}{4a\tau_{s}}\right),\label{eq:sol1}\\
\tau_{s}^{3/2} & \simeq & \frac{\hat{\xi}}{2}(1+\frac{1}{2a\tau_{s}}),\label{eq:sol2}\end{eqnarray}
where we've written $\hat{\xi}=(\frac{S+\bar{S}}{2})^{3/2}\xi$. Note
that extremizing with respect to $\tau_{s}$ gives us an exponentially
large volume and the three displayed terms in $V_{F}$ are all of
order ${\cal V}^{-3}$. This would mean that that at the classical
(negative) minimum found in \citep{Balasubramanian:2005zx}, the contribution
to the F-term potential from the dilaton and complex-structure moduli%
\footnote{At this point we ignore uplifting issues.%
} are zero. Also $V_{D}=0$ since it is positive definite and of order
$1/{\cal V}^{2}$. This would mean that, at least based on classical
considerations, $\tau_{a}\rightarrow0$, so that the standard model
cycle shrinks to zero or at least shrinks below the string scale.
Of course one might need to include all $\alpha'$ corrections in
such a case, but even so the important point here is that both the
D term and hence also the F-term of the MSSM cycle modulus $\tau_{a}$
become negligible at this classical minimum \citep{Blumenhagen:2009gk}.
We also note for future use that \eqref{eq:sol1} implies that\begin{equation}
a\tau_{s}=|\ln m_{3/2}|+O(1),\label{eq:ataus}\end{equation}
 and for $m_{3/2}\sim10-100TeV$ this is a number of $O(10)$.

The minimum found in \citep{Balasubramanian:2005zx} is at a negative
value of the (classical) cosmological constant\begin{equation}
V_{0}=-\frac{3\hat{\xi}}{16a\tau_{s}}\frac{m_{3/2}^{2}}{{\cal V}}.\label{eq:V0}\end{equation}
Note that here and in the rest of the paper we will be using the formula
for the gravitino mass\begin{equation}
m_{3/2}^{2}=e^{K}|W|^{2}\sim\frac{|W|^{2}}{{\cal V}^{2}}.\label{eq:gravitinomass}\end{equation}

\section{Classical soft terms}

Let us first compute the classical soft mass term in this model. The
general expression for the squared soft mass in SUGRA is \citep{Kaplunovsky:1993rd}\citep{Brignole:1997dp}\begin{eqnarray}
m_{\alpha\bar{\beta}}^{2} & = & V_{class}|_{0}\tilde{K}_{\alpha\bar{\beta}}+m_{3/2}^{2}\tilde{K}_{\alpha\bar{\beta}}-F^{A}F^{\bar{B}}R_{A\bar{B}\alpha\bar{\beta}}\label{eq:softmass1}\\
 & = & V_{class}|_{0}\tilde{K}_{\alpha\bar{\beta}}+m_{3/2}^{2}\tilde{K}_{\alpha\bar{\beta}}-F^{b}F^{\bar{b}}R_{b\bar{b}\alpha\bar{\beta}}\nonumber \\
 &  & -2ReF^{b}F^{\bar{s}}R_{b\bar{s}\alpha\bar{\beta}}-F^{s}F^{\bar{s}}R_{s\bar{s}\alpha\bar{\beta}}.\label{eq:softmass2}\end{eqnarray}
In the second equality we've used the fact that in the classical vacuum
before uplifting the only source of SUSY breaking are the F-terms
of $T^{b}$ and $T^{s}$. Also we will set $\tau_{a}\rightarrow0$
(and hence also set the corresponding F-term to zero) following the
arguments of \citep{Conlon:2008wa,Blumenhagen:2009gk}. To calculate
\eqref{eq:softmass2} explicitly we need these F-terms as well as
the matter metric $K_{\alpha\bar{\beta}}$. We find (see Appendix
A for details)

\begin{eqnarray}
F^{b} & = & -\tau^{b}\left(2+\frac{3}{8}\frac{\hat{\xi}}{a\tau^{s}}\frac{1}{{\cal V}}+O\left(\frac{1}{(a\tau^{s})^{2}{\cal V}}\right)\right)m_{3/2},\label{eq:Fb}\\
F^{s} & = & -\frac{3}{2}\frac{\tau^{s}}{a\tau^{s}}m_{3/2}(1+O({\cal V}^{-1})).\label{eq:Fs}\end{eqnarray}
 For the MSSM on D3 branes the matter metric can be calculated (see
Appendix A) from the formulae for the Kaehler coordinates given in
\citep{Grana:2003ek} (assuming that the formulae given in that reference
remain valid for D3 branes at a singularity). Since we expect D7 branes
on a collapsed cycle to act like D3 branes, these formulae should
be valid in that case too. We have \begin{equation}
K_{\alpha\bar{\beta}}=\frac{c}{{\cal V}+\hat{\xi}/2}(\sqrt{\tau^{b}}\omega_{\alpha\bar{\beta}}^{b}-\sqrt{\tau^{s}}\omega_{\alpha\bar{\beta}}^{s}),\label{eq:mattermetric}\end{equation}
where $\omega^{b(s)}$ is the harmonic $(1,1)$ form associated with
the big(small) modulus evaluated at the position of the D3 brane or
the collapsed cycle wrapped by the D7 brane, and $c$ is an $O(1)$
constant. 

Let us pause here for a moment to discuss the validity of (\ref{eq:mattermetric}),
beyond the context of smooth CY orientifolds within which it was derived
(see Appendix A) following the formula for the embedding of $ $D3
branes given in \citep{Grana:2003ek}. The basic argument in Appendix
A depended essentially on the field redefinition that is necessary
to obtain the correct holomorphic coordinates on moduli space, given
in the above reference. This is schematically of the form (see equation
(3.13) of \citep{Grana:2003ek})\begin{equation}
T^{i}+T^{\bar{i}}=2\tau^{i}+2\mu l^{2}i\omega_{\alpha\bar{\beta}}^{i}C^{\alpha}C^{\bar{\beta}}+(CUC+\bar{C}\bar{U}\bar{C}){\rm term}+\ldots\label{eq:Ttau0}\end{equation}
This formula was obtained by comparing the effective action of IIB
string theory with D3 branes located at some (generic smooth) point
on the CY orientifold with the standard supergravity formula for the
kinetic terms of the effective action which are written in terms of
a Kaehler potential. Now unfortunately a similar derivation has not
been given when the branes are located at a singularity. In so far
as one still expects a supergravity description of the low energy
physics, the question is to what extent a formula such as \eqref{eq:Ttau0}
remains valid. For us the essential feature of this formula is the
dependence on the matter fields to quadratic order. In particular
for the above calculation of the dependence of the metric on the Kaehler
moduli, what is relevant is the second term. So our assumption here
is that this structure, i.e. the proportionality to the $1,1$ form
$\omega^{i}$ (of the cycles which are stabilized) is preserved beyond
the original calculation at a generic smooth point. It is hard to
imagine that (apart from the possible modification of the coefficient)
that this structure can be qualitatively modified when the D3 brane
location is at some singularity. Indeed what we are using for our
calculations are just very generic features of the formula for the
metric \eqref{eq:mattermetric}. The numerical value of the normalized
diagonal mass term (see equation \eqref{eq:softmassclassical} will
certainly not change since it depends only on the part of the curvature
on moduli space that is proportional to the matter metric $K_{\alpha\bar{\beta}}$
and is independent on details of the dependence on the two $\omega$'s
. Furthermore the relative numerical coefficients in \eqref{eq:mattermetric}
(provided they are not changed by more than $O(1)$ numbers) is not
going to affect the qualitative features pertaining to the FCNC issue
discussed below either. In fact what is relevant for the latter is
the different dependence on the two moduli that is a feature of the
two terms in the metric, and this is unlikely to change for the metric
extracted from D3 branes at a singularity or D7 branes wrapping a
collapsing cycle.

The dependence on $\omega^{s}$ tends to give FCNC effects and we'll
postpone that discussion to the next section. Dropping that term the
metric is of the form $K_{\alpha\bar{\beta}}=f(\tau^{b},\tau^{s})\omega_{\alpha\bar{\beta}}^{b}$
and the Riemann tensor can be calculated from the formula $R_{i\bar{j}\alpha\bar{\beta}}=(\partial_{i}\partial_{\bar{j}}\ln f)K_{\alpha\bar{\beta}}$.
This gives (after using (\ref{eq:sol2}) and neglecting $O(1/(a\tau^{s})^{2}{\cal V}$)
terms %
\footnote{Typically $\hat{\xi}$ is a number of $O(1)$ and $a\tau^{s}$ is
a number around $30$ so these corrections can be safely ignored.
For instance even if $\hat{\xi}\sim5$ the ratio is expected to be
of around 15-20\% and that is the order of corrections that we would
expect to these formulae. This is certainly does not affect the qualitative
features of the calculations of the classical soft masses.%
})\begin{eqnarray}
R_{b\bar{b}\alpha\bar{\beta}} & = & \frac{1}{4(\tau^{b})^{2}}(1+\frac{15}{16}\frac{\hat{\xi}}{a\tau^{s}{\cal V}})K_{\alpha\bar{\beta}},\label{eq:Rbb}\\
R_{b\bar{s}\alpha\bar{\beta}} & = & -\frac{9}{16}\frac{(\tau^{s})^{1/2}}{(\tau^{b})^{5/2}}K_{\alpha\bar{\beta}},\label{eq:Rbs}\\
R_{s\bar{s}\alpha\bar{\beta}} & = & \frac{3}{16}\frac{(\tau^{s})^{-1/2}}{(\tau^{b})^{3/2}}K_{\alpha\bar{\beta}}.\label{eq:Rss}\end{eqnarray}
Using the above and equations (\ref{eq:Fb})\eqref{eq:Fs} in \eqref{eq:softmass2}
we find%
\footnote{To leading order in the matter fields $K_{\alpha\bar{\beta}}$ is
the same as $\tilde{K}_{\alpha\bar{\beta}}$ so we will not distinguish
between them in the rest of the paper.%
} to leading order in $1/a\tau_{s}{\cal V}$ (see Appendix A for details),\begin{equation}
m_{\alpha\bar{\beta}}^{2}=V_{0}K_{\alpha\bar{\beta}}+\frac{3}{8}\frac{\hat{\xi}}{a\tau^{s}}\frac{m_{3/2}^{2}}{{\cal V}}K_{\alpha\bar{\beta}}=+\frac{3}{16}\frac{\hat{\xi}}{a\tau^{s}}\frac{m_{3/2}^{2}}{{\cal V}}K_{\alpha\bar{\beta}}.\label{eq:softmassclassical}\end{equation}
Note that the second term differs in sign from the result quoted in
\citep{Blumenhagen:2009gk}. The reason is that we have here included
the contribution of the $F^{s}$ terms (and also the sub leading corrections
to $R_{b\bar{b}\alpha\bar{\beta}}$). Thus the classical squared masses
(even after including the contribution of the negative CC) is actually
positive.

\subsection{Uplift issues}

So far we have worked with the LVS minimum, which breaks supersymmetry,
but has a negative cosmological constant. It was argued in \citep{Balasubramanian:2005zx}
that at this minimum one expects the positive definite contributions
to the potential coming from the dilaton and the complex-structure
moduli to be actually zero since they scaled as ${\cal V}^{-2}$ whereas
the negative minimum scaled as ${\cal V}^{-3}$ . However this minimum
needs to be uplifted and one way that could happen is if the dilaton
($S$) and the complex structure moduli ($U^{m}$) acquired F-terms.
Generically one would expect at the new uplifted minimum (given that
$V_{0}\sim-m_{3/2}^{2}/({\cal V}\ln m_{3/2})$)

\begin{eqnarray}
F^{S}\bar{F}^{\bar{S}}K_{S\bar{S}} & \lesssim & \frac{m_{3/2}^{2}}{\ln m_{3/2}{\cal V}},\label{eq:FS}\\
F^{m}\bar{F}^{\bar{n}}K_{m\bar{n}} & \lesssim & \frac{m_{3/2}^{2}}{\ln m_{3/2}{\cal V}}.\label{eq:FU}\end{eqnarray}
Since much of the rest of the discussion focuses on the large modulus
we will often replace $T^{b}\rightarrow T$ and $\tau^{b}\rightarrow\tau$.
We have used above the argument that the seven brane when wrapping
a shrinking 4-cycle should behave like a D3 brane. For the same reason
we expect the leading classical contribution to the gauge coupling
function for both the D3 and D7 cases to be \begin{equation}
f_{a}=S+\kappa T^{a},\label{eq:fS}\end{equation}
(with $\kappa=0$ in the D3 case). Since as argued in \citep{Conlon:2008wa,Blumenhagen:2009gk}
the F-term of the modulus $T^{a}$ corresponding to the shrinking
cycle is vanishingly small, we see that without the uplift contribution
from the dilaton, there would be no (classical) contribution to the
gaugino mass in both the D3 and the D7 brane cases.

After uplift (assuming that $F^{S},F^{U}$take generic values consistent
with \eqref{eq:FS}\eqref{eq:FU}) we have the following expressions
for the gaugino mass $M$, the scalar mass $m$, the $A$ term, the
effective $\mu$ term %
\footnote{Assuming that the supersymmetric one is zero as is the case for the
D3 branes \citep{Grana:2003ek}.%
} and $B$ terms (see for example \citep{Kaplunovsky:1993rd} for definitions
and general formulae and Appendix A):\begin{eqnarray}
M_{a} & = & \frac{F^{i}\partial_{i}f_{a}}{2f_{a}}=\frac{F^{S}}{2S}\lesssim O\left(\frac{m_{3/2}}{\sqrt{\ln m_{3/2}{\cal V}}}\right),\label{eq:Mclassical}\\
m_{\alpha\bar{\beta}}^{2} & = & (m_{3/2}^{2}K_{\alpha\bar{\beta}}-F^{i}F^{\bar{j}}R_{i\bar{j}\alpha\bar{\beta}})=(O\left(\frac{m_{3/2}^{2}}{\ln m_{3/2}{\cal V}}\right))K_{\alpha\bar{\beta}}+\ldots,\label{eq:mclassical}\\
A_{\alpha\beta\gamma} & = & e^{K/2}F^{i}D_{i}y_{\alpha\beta\gamma}\lesssim O(\frac{m_{3/2}}{{\cal V}})y_{\alpha\beta\gamma},\label{eq:Aclassical}\\
\mu & \sim & B\mu/\mu\lesssim O\left(\sqrt{h_{21}}\frac{m_{3/2}}{\sqrt{\ln m_{3/2}{\cal V}}}\right).\label{eq:Bclassical}\end{eqnarray}
Note that in the last equation $h_{21}$ is the number of complex
structure moduli.

\subsection{Classical FCNC effects\label{sub:Classical-FCNC-effects}}

In computing the soft mass using \eqref{eq:mattermetric} we ignored
the second term inside the parenthesis. Let us now compute its contribution
(for details see Appendix A). To do so we must evaluate the Riemann
tensor. The leading contribution comes from the sectional curvature 

\begin{eqnarray}
R_{T\bar{T}\alpha\bar{\beta}} & = & \partial_{T}\partial_{\bar{T}}K_{\alpha\bar{\beta}}-K^{\gamma\bar{\delta}}\partial_{T}K_{\alpha\bar{\delta}}\partial_{\bar{T}}K_{\gamma\bar{\beta}}+O(C)\nonumber \\
 & = & \frac{1}{3}K_{T\bar{T}}c[\frac{\omega_{b}}{\tau_{b}}-\frac{7}{4}\frac{\omega_{s}}{\tau_{b}}\sqrt{\frac{\tau_{s}}{\tau_{b}}}]_{\alpha\bar{\beta}}+O(C).\label{eq:R}\end{eqnarray}
The problem is that this is not proportional to $K_{\alpha\bar{\beta}}$
- see \eqref{eq:mattermetric}. The $\omega^{s}$ dependence in \eqref{eq:R}
gives (from the third term on the RHS of \eqref{eq:softmass2}) an
additional term to the expression in \eqref{eq:softmassclassical},
i.e. \begin{eqnarray}
m_{\alpha\bar{\beta}}^{2} & = & m_{3/2}^{2}K_{\alpha\bar{\beta}}-F^{T}F^{\bar{T}}R_{T\bar{T}\alpha\bar{\beta}}+\ldots\nonumber \\
 & = & \frac{3}{16}\hat{\xi}\frac{m_{3/2}^{2}}{\ln m_{3/2}{\cal V}}K_{\alpha\bar{\beta}}+m_{3/2}^{2}\frac{3}{4}\sqrt{\frac{\tau_{s}}{\tau_{b}}}K'_{\alpha\bar{\beta}},\label{eq:softFCNC}\\
 & \equiv & (m^{2}\delta_{\alpha}^{\gamma}+\Delta m_{\alpha}^{2\gamma})K_{\gamma\bar{\beta}}.\end{eqnarray}
Here $K'_{\alpha\bar{\beta}}\equiv c\omega_{\alpha\bar{\beta}}^{s}/\tau_{b}$
and is not proportional to $K_{\alpha\bar{\beta}}$ so $\Delta m_{\alpha}^{2\gamma}=m_{3/2}^{2}\frac{3}{4}\sqrt{\frac{\tau_{s}}{\tau_{b}}}(K'K^{-1})_{\alpha}^{\gamma}$
is not a diagonal matrix. Also $m^{2}=\frac{3}{16}\hat{\xi}\frac{m_{3/2}^{2}}{\ln m_{3/2}{\cal V}}$
. If the two harmonic one-one forms $\omega^{b}$ and $\omega^{s}$
evaluated at the position of the D3 branes (or the collapsed cycle
wrapped by the D7 branes) are of the same order of magnitude then
the dominant contribution is a flavor violating one.

Clearly this would be a phenomenological disaster since at least for
the first two generations the flavor violating non-diagonal contributions
to the squared soft masses should be suppressed relative to the flavor
conserving ones. The relevant bound may be expressed by the following
relation (see for example \citep{Luty:2005sn}) \begin{equation}
\frac{\Delta m^{2}}{m^{2}}\lesssim10^{-3}\frac{m}{500GeV}.\label{eq:phenoconstraint}\end{equation}
 So if we want soft masses $m\lesssim1TeV$ in order to address the
hierarchy problem the FCNC effect must be suppressed by at least a
factor $10^{-3}$. Then the following alternatives may be pursued.

\begin{itemize}
\item To be consistent with \eqref{eq:phenoconstraint} we need $\omega^{s}\lesssim10^{-3}\frac{1}{\ln m_{3/2}\tau^{b}}\omega^{b}$
at the MSSM point. We can get this as follows. The small cycle may
be regarded as a blow up (by a $P^{2}$) of a singularity. Then (at
least in a non-compact Calabi-Yau) it has been shown in \citep{Lutken:1987ny}
that the corresponding harmonic 1,1 form falls off as $R^{-6}$ where
$R$ is the distance to the singularity%
\footnote{We thank Joe Conlon for suggesting this and bringing reference \citep{Lutken:1987ny}
to our attention.%
}. It is quite plausible that this behavior applies to the small cycle
of a compact Calabi-Yau. Then indeed the desired suppression can be
obtained if the distance $R$ is identified with the location of the
$D3$ brane (or the MSSM collapsed cycle) and is of the order of ${\cal V}^{1/6}\sim\tau^{1/4}$.
The FCNC suppression is then obtained provided ${\cal V}\gtrsim10^{12}$.
This would imply a string scale that is well below the Planck scale
since $M_{string}\sim M_{P}/\sqrt{{\cal V}}\lesssim10^{12}GeV$$ $.
Any hope of getting a GUT scenario within LVS is then eliminated,
but we would still have a viable intermediate scale phenomenology.
This of course is consistent with the usual LVS scenario as discussed
in \citep{Conlon:2006wz} and references therein. Nevertheless it
should be stressed that this conclusion holds only if we ignore quantum
- in particular Weyl anomaly and gaugino mediation - contributions
to scalar masses (see next section).
\item The alternative within LVS is to have a very heavy soft mass scale
$\gtrsim10^{3}TeV$. In this case the SUSY solution to the hierarchy
problem is much more fine-tuned than in the previous one. Nevertheless
it is interesting to note that the constraint on the volume is now
much weaker and reads ${\cal V}\gtrsim10^{3}$. Note that in this
case one might hope that the gaugino masses are still at the TeV scale
(this could have been the case according to \eqref{eq:Mclassical}
if the dilaton $ $F-term is not responsible for the uplift and the
gaugino mass arises from loop corrections to the gauge coupling function).
This scenario is essentially that of split supersymmetry. However
again the Weyl anomaly and gaugino mediation effects will modify this.
\item The third possibility is to consider compactifications with just one
Kaehler modulus. In this case an LVS solution is not possible. But
one could by including the $\alpha'$ corrections and race track terms
find an intermediate volume (${\cal V}\sim10^{3-4}$) solution. Of
course in this case $W_{0}$ the flux superpotential would have to
be fine tuned to extremely low values in order to get TeV scale soft
masses. This scenario has been discussed in \citep{deAlwis:2008kt}.
\end{itemize}
In the following we will pursue only the first of these alternatives.

\section{Quantum Effects}

\subsection{String loop effects}

String loop contributions to the classical contributions considered
in the previous section can be estimated. This can be done either
from an effective field theory calculation as in \citep{Choi:1997de,deAlwis:2008kt}
or from arguments based on calculations in toy models in string theory
\citep{Blumenhagen:2009gk}. They agree if certain cancellations take
place. This is essential if the original LVS minimum is not to be
destabilized. In this case these do not give a significant correction
to the classical soft terms discussed above.

\subsection{Weyl Anomaly and Gaugino masses}

Significant corrections to the soft terms can arise from Weyl anomaly
contributions \citep{Kaplunovsky:1994fg,Randall:1998uk,Giudice:1998xp,Bagger:1999rd,Dine:2007me,deAlwis:2008aq}.
These are independent of the size of the compactification once the
value of the gravitino mass is chosen. In particular the gaugino masses
are given by the expressions (see Appendix B for a discussion)\begin{equation}
M_{a}=-b_{a}\left(\frac{\alpha_{a}}{4\pi}\right)m_{3/2}.\label{eq:AMSBgaugino}\end{equation}
Here $a=1,2,3$ index the three standard model gauge groups - respectively
$U(1),\, SU(2),\, SU(3)$, with couplings $\alpha_{a}=g_{a}^{2}/4\pi$.
 These expressions when evaluated at the UV scale (assumed to be at
or close to the unification scale so that $\alpha_{a}\sim\alpha_{GUT}\sim1/25$)
give \begin{equation}
M_{1}=\frac{33}{5}\frac{\alpha_{GUT}}{4\pi}m_{3/2},\, M_{2}=\frac{\alpha_{GUT}}{4\pi}m_{3/2},\, M_{3}=-3\frac{\alpha_{GUT}}{4\pi}m_{3/2}.\label{eq:gauginomass}\end{equation}
These should be treated as initial values for the RG evolution down
to the MSSM scale - and numerically have values $O(10^{-2}-10^{-3})m_{3/2}$.
The important point here is that these values are larger than the
classical contribution which gave gaugino masses $\lesssim O(m_{3/2}/\ln m_{3/2}\sqrt{{\cal V}})$
(see \eqref{eq:Mclassical}), unless ${\cal V}\lesssim10^{2}$ which
is far too small a volume for an LVS scenario because of the FCNC
problem.

\subsection{Gaugino Mediation}

Now using the above values as boundary conditions for the RG evolution
down to the MSSM scale scalar masses are generated by the gaugino
mediation  mechanism \citep{Kaplan:1999ac,Chacko:1999mi}. As shown
in Appendix B we get,

\begin{equation}
m_{1}^{2}\sim m_{2}^{2}\sim10^{-6}m_{3/2}^{2},\, m_{3}^{2}\sim10^{-4}m_{3/2}^{2}.\label{eq:AMSBscalar2}\end{equation}
This is to be compared to the (diagonal) classical contribution \eqref{eq:softFCNC}
$m^{2}\sim m_{3/2}^{2}1/(\ln m_{3/2}{\cal V})$ which would dominate
over (\ref{eq:AMSBscalar2}) only if ${\cal V}\lesssim10^{3}$. But
in that case the flavor non-diagonal contribution (see \eqref{eq:softFCNC})
would be far too large (even assuming the suppression by a factor
${\cal V}$ of the 1,1 form $\omega^{s}$ that we argued for in the
discussion after \eqref{eq:softFCNC}). 

Thus we require that \eqref{eq:AMSBscalar2} dominates the FCNC contribution
in \eqref{eq:softFCNC} which is $\Delta m^{2}\sim K'K^{-1}m_{3/2}^{2}/\sqrt{\tau^{b}}\sim m_{3/2}^{2}/(\tau^{b})^{2}$
(see \eqref{eq:softFCNC}\eqref{eq:phenoconstraint}) , by at least
a factor of $10^{3}$. i.e. we need to have \begin{equation}
\frac{\Delta m^{2}}{m_{3}^{2}}\sim\frac{10^{4}}{(\tau^{b})^{2}}\lesssim10^{-3}.\label{eq:FCNC2}\end{equation}
This gives \[
{\cal V}\sim(\tau^{b})^{3/2}\gtrsim10^{5}.\]
This would yield an effective string scale of $M_{string}\lesssim1/\sqrt{{\cal V}}\sim10^{-2.5}M_{P}\sim10^{15.5}GeV$
which may just accommodate a GUT scenario. 

Be that as it may, to have a SUSY solution to the hierarchy problem
in a GUT scenario clearly needs fine tuning of the flux superpotential.
From \eqref{eq:AMSBscalar2} we see that to get $TeV$ scale squark
masses $m_{3}\sim1TeV$, the gravitino mass must be $m_{3/2}\sim\frac{|W|}{{\cal V}}\sim10^{2}TeV$.
For $W\sim O(1)$ this gives ${\cal V}\sim10^{13}$ (well above our
lower limit) and a string scale $M_{string}\sim M_{P}/\sqrt{{\cal V}}\sim10^{12}$
which is certainly well below the GUT scale. To get a scale close
to the GUT scale would need a highly fine-tuned $W\sim10^{-8}$.

\section{Conclusions\label{sec:Conclusions}}

In this paper we have discussed type IIB compactifications on ``Swiss
Cheese'' type manifolds with the MSSM either on a D3 brane(s) at
a singularity or on a stack of D7 branes which wrap a 4-cycle. The
conflict between chirality and the generation of a non-perturbative
superpotential leads to the conclusion that the latter (MSSM) cycle
actually collapses below the string scale (or to a singularity) and
its modulus does not contribute to SUSY breaking, so that effectively
one can ignore it. The following results were derived in this set
up.

\begin{enumerate}
\item The (classical) diagonal soft mass is given by $m_{soft}=\sqrt{\frac{3\hat{\xi}}{16\ln m_{3/2}{\cal V}}}m_{3/2}$.
However there is also a flavor violating contribution to the mass
matrix which needs to be suppressed. Assuming that the 1,1 form associated
with the small cycle falls off with distance, as in the non-compact
case studied in \citep{Lutken:1987ny}, we get a lower bound ${\cal V}\gtrsim10^{12}$
in order that there is sufficient suppression of FCNC effects relative
to the classical soft mass. 
\item However the Weyl anomaly contribution to the gaugino mass actually
dominates the classical contribution. Furthermore this generates,
through the mechanism of gaugino mediation, a quantum contribution
at the MSSM scale to the scalar soft masses, that is also larger than
the classical contribution. Hence the scenario of item 1. above gets
modified. These LVS models appear to give a string theoretic construction
of a sequestered situation as envisaged in \citep{Randall:1998uk}%
\footnote{To truly establish sequestering one would of course need to demonstrate
that the couplings of the bulk and brane fields are suppressed at
the quantum level as well. This would require a calculation of string
loop effects beyond that considered in the literature. For some conjectures
regarding this see \citep{Conlon:2008wa} and references therein.
It should also be noted that these assumptions are in agreement with
the effective field theory estimates given in \citep{deAlwis:2008kt}.
Essentially the point is that the quantum corrections are also expected
to be suppressed by large volume factors, so that it seems unlikely
that these corrections could significantly change the classical results.
However it would be nice if this could be supported by a detailed
calculation.%
}. The suppression of FCNC effects now only gives the weaker constraint
${\cal V}\gtrsim10^{5}$ (see discussion after \eqref{eq:FCNC2}).
\item The gravitino mass in this scenario is $m_{3/2}\sim10^{2}TeV$ if
we wish to have TeV scale SUSY breaking MSSM soft terms. The gravitino
gives no cosmological problems but the sGoldstino (light modulus)
mass would be $m_{mod}\sim m_{3/2}/\sqrt{{\cal V}}<1TeV$ so this
scenario appears to suffer from the cosmological modulus problem.
Again the lower bound on ${\cal V}$ is compatible with this estimate
of the gravitino mass only if $W_{0}$ is highly fine-tuned to values
around $10^{-7}$. If $W_{0}\sim O(1)$ then ${\cal V}\sim10^{13}$
, we have an intermediate string scale, no possibility of Grand Unification,
and a very light modulus $\sim100MeV$! Furthermore there would be
a serious $\mu$ problem since as we see from \eqref{eq:Bclassical}
$\mu$ is highly suppressed for large volumes.
\item The resolution of the cosmological modulus problem necessitates raising
the gravitino mass to $m_{3/2}\sim10^{3}TeV$. However the Weyl anomaly
generated gaugino mass is now $\sim10TeV$ with gaugino mediated soft
masses which are at least a few TeV. We may take the volume ${\cal V}$
close to its minimum possible value $10^{3}$, consistent with the
now somewhat less stringent FCNC constraint \eqref{eq:phenoconstraint}
(since the squark mass is higher). Then we have a somewhat less fine-tuned
$W_{0}\sim10^{-6}$ and a light modulus $\sim10TeV$. But of course
addressing the hierarchy problem would require somewhat more fine-tuning.
If the cosmological modulus problem is taken seriously then this should
be considered the preferred solution within this class of models. 
\end{enumerate}
The LVS compactification of type IIB string theory thus gives us an
appealing class of models of supersymmetry breaking and transmission.
It is consistent with all theoretical constraints and satisfies phenomenological
constraints (and in the case of 4 above cosmological ones as well)
. It is predictive and is expected to be ultra-violet complete. As
argued in the introduction a bottom up approach ignores the necessary
embedding of the low energy (or intermediate scale) hidden sector
dynamics in the larger framework of a string theory. The essential
point here is that one cannot ignore the dynamics of the string theory
moduli and focus on some additional sector (as is usually done in
GMSB) since in SUGRA, both open string fields and closed string moduli
are coupled together in a highly non-linear fashion. This class of
models, with LVS compactifications, generically break supersymmetry
and provide, within a string theory context, a very compelling and
phenomenologically viable scenario in which to discuss MSSM SUSY breaking.
The detailed phenomenology of these models will be discussed elsewhere
\citep{Baer:2009tp}.

\section{Acknowledgements}

I'm very grateful to Fernando Quevedo for collaboration in the early
stages of this work and for extensive discussions on LVS models. Special
thanks are also due to Joe Conlon for discussions and for drawing
my attention to \citep{Lutken:1987ny}. I also wish to acknowledge
discussions with Cliff Burgess, Oliver DeWolfe and Matthew Headrick
. This research is supported in part by the United States Department
of Energy under grant DE-FG02-91-ER-40672.

\section*{Appendix A}

We give here some details of the calculation of the F-terms of the
two Kaehler moduli $\tau^{b},\tau^{s}$. Useful formulae for calculating
the metrics inverse metrics and Riemann tensors for Kaehler potentials
of the form $K=-n\ln Y$are given in Appendix A of \citep{Covi:2008ea}.
In our case $n=2$ and $Y={\cal V}+\frac{\hat{\xi}}{2}$ where ${\cal V}=(\tau^{b})^{3/2}-(\tau^{s})^{3/2}$
where $\tau^{i}=\frac{1}{2}(T^{i}+\bar{T}^{\bar{i}})$ with $T^{i}$
being the holomorphic Kaehler moduli. As usual $K_{i}\equiv\partial_{T^{i}}K,$
etc. We find \begin{equation}
K^{i\bar{j}}K_{\bar{j}}\sim-2\tau^{i}-\frac{3}{2}\hat{\xi}\frac{\tau^{i}}{{\cal V}},\label{eq:KijbarKjbar}\end{equation}
 and \begin{eqnarray}
K^{s\bar{s}} & = & -2({\cal V}+\frac{\hat{\xi}}{2})(-\frac{4}{3}(\tau^{s})^{1/2}+4(\tau^{s})^{2}+O({\cal V}^{-1})),\label{eq:Kssbar}\\
K^{b\bar{s}} & = & 4\tau^{b}\tau^{s}(1+O({\cal V}^{-1})).\label{eq:Kbsbar}\end{eqnarray}
 Also note that the stabilization of the axion corresponding to the
small modulus results in a sign flip (i.e. the axion takes a value
that is an odd multiple of $\pi$ after choosing without loss of generality
the phases of $W_{0},A$ to be zero) so that effectively the superpotential
is $W=W_{0}+Ae^{-aT^{s}}=W_{0}-Ae^{-a\tau^{s}}$ \citep{Balasubramanian:2005zx}.
Then we find (note that $F^{i}\equiv e^{K/2}K^{i\bar{j}}(\partial_{\bar{j}}\bar{W}+K_{\bar{j}}\bar{W})$)
using the solution \eqref{eq:sol1}\eqref{eq:sol2}, \begin{equation}
F^{b}=-\tau^{b}(2+\frac{3}{2}.\frac{\hat{\xi}}{4a\tau^{s}}\frac{1}{{\cal V}})m_{3/2},\label{eq:Fbapp}\end{equation}
\begin{equation}
F^{s}=-\frac{3\tau^{s}}{2a\tau^{s}}m_{3/2}(1+O({\cal V}^{-1})).\label{eq:Fsapp}\end{equation}
($F^{b}$ was also calculated in \citep{Blumenhagen:2009gk}).

Here we compute the matter metric for the matter located on a D3 brane
which sits at a point in the internal Calabi-Yau space. We expect
that a similar formula will be valid for matter on a stack of D7 branes
wrapping a collapsing four cycle.

The holomorphic Kaehler moduli $T^{i}$ are related to the moduli
$\tau^{i}$ (in terms of which the volume is given by \eqref{eq:Swisscheese})
by (see equation (3.13) of \citep{Grana:2003ek})\begin{equation}
T^{i}+T^{\bar{i}}=2\tau^{i}+2\mu l^{2}i\omega_{\alpha\bar{\beta}}^{i}C^{\alpha}C^{\bar{\beta}}+(CUC+\bar{C}\bar{U}\bar{C}){\rm term}+\ldots\label{eq:Ttau}\end{equation}
to linear order in the complex structure moduli $U$. Here $\omega^{i}$
are the harmonic 1,1 forms on the CY orientifold evaluated at the
position of the D3 brane, $C^{\alpha}$ are the matter fields on the
D3 brane, $l$ is the axionic partner of the dilaton ($S\equiv e^{-\phi}-il$),
and $\mu$ is the tension of the D3 brane. Writing ${\cal V}_{i}\equiv\partial{\cal V}/\partial\tau^{i}$,\[
K_{\alpha}\equiv\frac{\partial K}{\partial C^{\alpha}}|_{T,U,S}=-2\frac{{\cal V}_{i}}{Y}\frac{\partial\tau^{i}}{\partial C^{\alpha}}|_{T,U,S}\]
Differentiating \eqref{eq:Ttau} with respect to $C$ keeping the
moduli $T,U,S$ fixed we have \begin{eqnarray*}
0 & = & 2\frac{\partial\tau^{i}}{\partial C^{\alpha}}+2\mu(i\omega_{\alpha\bar{\beta}}^{i})C^{\bar{\beta}}+O(UC)\\
0 & = & 2\frac{\partial^{2}\tau^{i}}{\partial C^{\alpha}\partial C^{\bar{\beta}}}+2\mu(i\omega_{\alpha\bar{\beta}}^{i})\end{eqnarray*}
Hence we have\begin{equation}
K_{\alpha\bar{\beta}}=2\mu i\omega_{\alpha\bar{\beta}}^{i}\frac{{\cal V}_{i}}{Y}+O(C^{2})=\frac{3\mu}{Y}(i\omega_{\alpha\bar{\beta}}^{b}\sqrt{\tau^{b}}-i\omega_{\alpha\bar{\beta}}^{s}\sqrt{\tau^{s}})+O(C^{2}),\label{eq:mattmetapp}\end{equation}
where in the last step we specialized to the Swiss cheese CY manifold
\eqref{eq:Swisscheese} (with $\tau^{a}\rightarrow0$). Note also
for future reference that \begin{equation}
Z=K_{H_{1}H_{2}}\sim\frac{{\cal V}_{i}}{Y}\frac{\partial^{2}\tau^{i}}{\partial H_{1}\partial H_{2}}.\label{eq:Z}\end{equation}

To compute the soft masses (and other soft terms) we need the sectional
curvatures.

First let us ignore the second term in parenthesis in \eqref{eq:mattmetapp}.
In this case the matter metric is conformal to the 1,1 harmonic form
of the large modulus $K_{\alpha\bar{\beta}}=f(\tau^{b},\tau^{s})i\omega_{\alpha\bar{\beta}}^{b},\, f\equiv3\mu\sqrt{\tau^{b}}/Y$
and the relevant components of the Riemann tensor are easily computed
using the formula $R_{i\bar{j}\alpha\bar{\beta}}=\partial_{i}\partial_{\bar{j}}\ln fK_{\alpha\bar{\beta}}$.
This gives \begin{eqnarray}
R_{b\bar{b}\alpha\bar{\beta}} & = & \frac{1}{4(\tau^{b})^{2}}(1+\frac{15}{16}\frac{\hat{\xi}}{a\tau^{s}{\cal V}})K_{\alpha\bar{\beta}},\label{eq:Rbbapp}\\
R_{b\bar{s}\alpha\bar{\beta}} & = & -\frac{9}{16}\frac{(\tau^{s})^{1/2}}{(\tau^{b})^{5/2}}K_{\alpha\bar{\beta}},\label{eq:Rbsapp}\\
R_{s\bar{s}\alpha\bar{\beta}} & = & \frac{3}{16}\frac{(\tau^{s})^{-1/2}}{(\tau^{b})^{3/2}}K_{\alpha\bar{\beta}}.\label{eq:Rssapp}\end{eqnarray}
For future use we will recalculate $R_{b\bar{b}\alpha\bar{\beta}}$
keeping both terms in the parenthesis in \eqref{eq:mattmetapp} (but
ignoring the $\hat{\xi}$ dependence for simplicity), and using the
general formula for the Riemann tensor in Kaehler geometry\begin{eqnarray}
R_{b\bar{b}\alpha\bar{\beta}} & = & \partial_{b}\partial_{\bar{b}}K_{\alpha\bar{\beta}}-K^{\gamma\bar{\delta}}\partial_{\bar{b}}K_{\alpha\bar{\delta}}\partial_{b}K_{\gamma\bar{\beta}},\nonumber \\
 & = & \frac{3\mu}{4(\tau^{b})^{3}}(i\omega_{\alpha\bar{\beta}}^{b}-\frac{7}{4}\sqrt{\frac{\tau^{s}}{\tau^{b}}}i\omega_{\alpha\bar{\beta}}^{s}),\label{eq:Rbbnew}\\
 & = & \frac{1}{3}K_{b\bar{b}}(K_{\alpha\bar{\beta}}-K'_{\alpha\bar{\beta}}\sqrt{\frac{\tau^{s}}{\tau^{b}}}).\label{eq:Rbbnew2}\end{eqnarray}
Here we have defined \[
K'_{\alpha\bar{\beta}}\equiv\frac{9\mu}{4}\frac{i\omega_{\alpha\bar{\beta}}^{s}}{\tau^{b}},\]
 to be compared with $K_{\alpha\bar{\beta}}\sim\frac{3\mu}{\tau^{b}}i\omega_{\alpha\bar{\beta}}^{b}$. 

Let us use the above results to calculate soft masses. These are given
by\begin{eqnarray*}
m_{\alpha\bar{\beta}}^{2} & = & V_{class}|_{0}\tilde{K}_{\alpha\bar{\beta}}+m_{3/2}^{2}\tilde{K}_{\alpha\bar{\beta}}-F^{b}F^{\bar{b}}R_{b\bar{b}\alpha\bar{\beta}}\\
 &  & -2ReF^{b}F^{\bar{s}}R_{b\bar{s}\alpha\bar{\beta}}-F^{s}F^{\bar{s}}R_{s\bar{s}\alpha\bar{\beta}}.\end{eqnarray*}
First let us compute the flavor diagonal part - i.e. we will ignore
the contribution to the matter metric from the harmonic form $\omega^{s}$.
We have \begin{eqnarray*}
F^{b}\bar{F}^{\bar{b}}R_{b\bar{b}\alpha\bar{\beta}} & = & m_{3/2}^{2}(1+\frac{21}{16}\frac{\hat{\xi}}{a\tau^{s}{\cal V}})K_{\alpha\bar{\beta}}\\
2F^{b}\bar{F}^{\bar{s}}R_{b\bar{s}\alpha\bar{\beta}} & = & -\frac{27}{8}\frac{\hat{\xi}}{2a\tau^{s}{\cal V}}m_{3/2}^{2}K_{\alpha\bar{\beta}}\\
F^{s}\bar{F}^{\bar{s}}R_{s\bar{s}\alpha\bar{\beta}} & = & \frac{27}{64}\frac{\hat{\xi}}{2(a\tau^{s})^{2}{\cal V}}m_{3/2}^{2}K_{\alpha\bar{\beta}}\end{eqnarray*}
So we have\[
m_{\alpha\bar{\beta}}^{2}=(V_{class}|_{0}+\frac{3}{8}\frac{\hat{\xi}}{a\tau^{s}{\cal V}}m_{3/2}^{2})\tilde{K}_{\alpha\bar{\beta}}+{\rm flavor\, non-diagonal}\]
Now let us compute the flavor non-diagonal piece. The leading contribution
comes from the extra contribution to $F^{b}\bar{F}^{\bar{b}}R_{b\bar{b}\alpha\bar{\beta}}$
coming from the term proportional to $K'_{\alpha\bar{\beta}}$ in
the expression for the Riemann tensor \eqref{eq:Rbbnew2}. Using $F^{b}\bar{F}^{\bar{b}}K_{b\bar{b}}\sim3m_{3/2}^{2}$
we find this to be\[
\Delta m_{\alpha\bar{\beta}}^{2}=\frac{3}{4}\sqrt{\frac{\tau^{s}}{\tau^{b}}}m_{3/2}^{2}K'_{\alpha\bar{\beta}}.\]
So unless $K'_{\alpha\bar{\beta}}$ is strongly suppressed relative
to $ $$\tilde{K}_{\alpha\bar{\beta}}$ (which means in effect the
suppression at the position of the D3brane(s) of $\omega^{s}$ compared
to $\omega^{b}$) there would be a serious FCNC problem. 

There is also an issue with the $\mu$ term (see \eqref{eq:Bclassical})
that needs to be discussed. If the uplift comes mainly from giving
F-terms to the complex structure moduli then from \eqref{eq:FU} it
follows that the $|F_{m}|\lesssim m_{3/2}/(\sqrt{h_{21}}\sqrt{\ln m_{3/2}{\cal V}})$
since in a basis in which the relevant metric is diagonal there are
effectively $h_{21}$ terms in the sum. Now the expression for the
effective $\mu$ term has a contribution $F^{m}\partial_{m}Z$. From
\eqref{eq:Z} we see that this gives the factor $\sqrt{h_{21}}$ in
the upper bound \eqref{eq:Bclassical}.

\section*{Appendix B}

The physical gauge coupling (at the cutoff scale/GUT scale $\Lambda$)
can be written in the following form \citep{Kaplunovsky:1994fg}\citep{ArkaniHamed:1997mj}\citep{deAlwis:2008aq} 

\begin{equation}
H_{i}=f_{i}-\frac{3c_{i}}{8\pi^{2}}\tau-\sum_{r}\frac{T_{i}(r)}{4\pi^{2}}\tau_{r}-\frac{T(G_{i})}{4\pi^{2}}\tau_{i}.\label{eq:Hphys}\end{equation}
Here $G_{i}$ is the gauge group and $c_{i}=T(G_{i})-\sum_{r}T_{i}(r)$
\footnote{Note that this is the negative of the coefficient defined in \citep{Kaplunovsky:1994fg}.%
} where $T(G_{i}),\, T_{i}(r)$ are respectively the trace of a squared
generator in the adjoint and the matter representations of the group.
The first term is the classical gauge coupling (say at the scale $\Lambda$).
The second arises from the Weyl anomaly which arises when one does
a chiral rotation from the supergravity frame to the Kaehler-Einstein
frame. The third term comes from an anomaly associated with the field
redefinition $C_{\alpha}\rightarrow e^{\tau_{Z}}C_{\alpha}$ needed
to get canonical normalization of the MSSM fields and the last term
arises from the redefinition of the gauge field pre-potential $V_{i}\rightarrow e^{(\tau_{i}+\bar{\tau}_{i})/2}V_{i}$.
The chiral fields $\tau,\tau_{r},\tau_{i}$ are fixed by the relations\begin{eqnarray}
\tau+\bar{\tau} & = & \frac{1}{3}K|_{{\rm harm}},\label{eq:phi}\\
\tau_{r}+\bar{\tau}_{r} & = & \ln\det\tilde{K}_{\alpha\bar{\beta}}^{(r)},\label{eq:tauZ}\\
\exp[-(\tau_{i}+\bar{\tau}_{i})]|_{{\rm harm}} & = & \frac{1}{2}(H_{i}+\bar{H}_{i}),\label{eq:tauV}\end{eqnarray}
It should be emphasized that (as observed by Kaplunovsky and Louis)
the first relation is precisely the one which accomplishes the transformations
and field redefinitions that are needed to get to the Einstein-Kaehler
frame. These are the same transformations that are done in for example
Wess and Bagger \citep{Wess:1992cp} in component form to get to the
final SUGRA action displayed in Appendix G of that work. The last
relation comes from supersymmetrizing the lowest component relation
$\exp[-(\tau_{i}+\bar{\tau}_{i})]|_{0}=1/g_{phys}^{(i)2}\equiv\Re H_{i}|_{0}$.
To get the physical coupling function at a low scale $\mu$ ( $\ll\Lambda$)
we need to evaluate the right hand sides of \eqref{eq:phi}\eqref{eq:tauZ}\eqref{eq:tauV}
at the scale $\mu$ and make the replacement\begin{equation}
f_{i}\rightarrow f_{i}-\frac{b_{i}}{16\pi^{2}}\ln\frac{\Lambda}{\mu}.\label{eq:fmu}\end{equation}
Here $b_{i}=3T(G_{i})-\sum_{r}T_{i}(r)$ and we used the fact that
$f_{i}$ is only renormalized at one-loop.

Projecting the lowest component of \eqref{eq:Hphys} (with the replacement\eqref{eq:fmu})
and using the relations \eqref{eq:phi}\eqref{eq:tauZ}\eqref{eq:tauV}
gives us (the integrated form of) the NSVZ relation (with SUGRA and
Weyl anomaly corrections)\begin{eqnarray}
\frac{1}{g_{{\rm phys}}^{(i)2}} & = & \Re f_{i}-\frac{b_{i}}{16\pi^{2}}\ln\frac{\Lambda}{\mu}-\frac{c_{i}}{16\pi^{2}}K|_{0}\nonumber \\
 &  & -\sum_{r}\frac{T_{i}(r)}{8\pi^{2}}\ln\det\tilde{K}_{\alpha\bar{\beta}}^{(r)}|_{0}+\frac{T(G_{i})}{8\pi^{2}}\ln\frac{1}{g_{{\rm phys}}^{(i)2}}.\label{eq:gphys}\end{eqnarray}
On the other hand projecting the F-term of \eqref{eq:Hphys} gives
(after solving for $M_{i}/g_{{\rm phys}}^{(i)2}$)\begin{eqnarray}
\frac{M_{i}}{g_{{\rm phys}}^{(i)2}} & = & \frac{1}{2}(F^{A}\partial_{A}f_{i}-\frac{c_{i}}{8\pi^{2}}F^{A}K_{A}-\sum_{r}\frac{T_{i}(r)}{4\pi^{2}}F^{A}\partial_{A}\ln\det\tilde{K}_{\alpha\bar{\beta}}^{(r)})\nonumber \\
 &  & \times(1-\frac{T(G_{i})}{8\pi^{2}}g_{{\rm {\rm phys}}}^{(i)2})^{-1}.\label{eq:m/g2}\end{eqnarray}
Let us evaluate this for the theory described in this paper. Since
the gauge coupling function $f_{a}=S$ whose F-term is highly suppressed,
the classical contribution can be ignored compared to the AMSB one
coming from the last two terms in the first parenthesis. To one-loop
(keeping only the large modulus ($T^{b}\equiv T$) contribution since
the other contributions are highly suppressed) we have \begin{equation}
M_{i}=-b_{i}\frac{g^{(i)2}}{16\pi^{2}}m_{3/2}=-b_{i}\frac{\alpha_{i}}{4\pi}m_{3/2}\label{eq:gauginomass'}\end{equation}
To get this we used $F^{T}=-(T+\bar{T})m_{3/2}$, $K_{T}=-3/(T+\bar{T})$
and $\tilde{K}_{\alpha\bar{\beta}}=k_{\alpha\beta}/(T+\bar{T}$).
Also note that there is no direct AMSB contribution to the scalar
masses since the Weyl anomaly just affects the gauge coupling function
- at least at the two derivative level %
\footnote{At the four derivative level of course there are curvature squared
contributions.%
}.

Nevertheless there is a contribution to the scalar masses coming from
`gaugino mediation' (see \citep{Luty:2005sn} for a review). This
comes about because of the contribution of gauginos to the running
of the scalar masses. The relevant equation is (see section 11 of
\citep{Luty:2005sn}) assuming GUT unification of coupling constants
($t\equiv\ln\mu/\Lambda$ and $c(r)$ is the quadratic Casimir in
$r$)\begin{equation}
\frac{dm_{scalar}^{2}}{dt}=-\frac{c(r)}{2\pi^{2}}g_{i}^{2}M_{i}^{2},\label{eq:massRG}\end{equation}
Integrating this using the beta function equations and the RG invariance
of $M_{i}/g_{i}^{2}$,\begin{equation}
m_{scalar}^{2}=\frac{2c(r)}{b_{i}}[\frac{g_{i}^{4}(\mu)}{g_{i}^{4}(\Lambda)}-1]M_{i}^{2}\simeq2c(r)\alpha_{GUT}\ln\frac{\Lambda}{\mu}M_{i}^{2}.\label{eq:mscalargaugino}\end{equation}
Note that in \eqref{eq:massRG}we have only kept the dominant terms.
Also we note that the squared scalar masses generated by this mechanism
are always positive. Furthermore for $\mu$ at the TeV scale and $\Lambda$
at the GUT scale and $\alpha_{GUT}\sim1/25$, the scalar mass is of
the order of the gaugino mass which in turn has a magnitude $m_{3}\sim10^{-2}m_{3/2}$,
$m_{2}\sim m_{1}\sim10^{-3}m_{3/2}$. Thus with $m_{3/2}\sim100TeV$
we have TeV scale gluino and squark masses. These clearly dominate
the classical masses that we obtained even in the most favorable case
with ${\cal V}\sim10^{6}$. These numbers would be somewhat different
in the intermediate string scale case, but the above mechanism i.e.
RG running from this intermediate scale, will still be the dominant
mechanism for generating the scalar masses and the A-term. A detailed
account of the phenomenology in both cases will be presented in \citep{Baer:2009tp}.
\[
\]

\bibliographystyle{apsrev}
\bibliography{myrefs}

\end{document}